\documentclass[preprint,eqsecnum,nofootinbib,natbib,aps]{revtex4}
\usepackage{graphicx}

\begin{document}
\title{Method for Solving the Bloch Equation
from the Connection with Time-Dependent Oscillator}
\author{Hyung Koo Kim}
\author{Sang Pyo Kim}
\email{sangkim@kunsan.ac.kr} \affiliation{Department of Physics,
Kunsan National University, Kunsan 573-701, Korea}

\date{\today}
\begin{abstract}
We introduce a novel method to find exact density operators for a
spin-1/2 particle in time-dependent magnetic fields by using the
one-mode bosonic representation of $su(2)$ and the connection with
a time-dependent oscillator. As illustrative examples, we apply
the method to find the density operators for constant and/or
oscillating magnetic fields, which turn out to be time-dependent
in general.
\end{abstract}

\maketitle

\section{Introduction}

Recently two-level atoms have been investigated under the
influence of pulsed or oscillating fields \cite{friedberg,allen}.
However, a system of spin-1/2 particles under external magnetic
fields has been a subject of interest since the advent of quantum
mechanics \cite{bloch}. Also, two-level systems of electric
dipoles under external electric fields have been studied for a
long time \cite{autler}, and the statistics of a spin-1/2 particle
has been found in terms of density matrices \cite{bloch,rabi}. All
these systems can be described by the Lie algebra $su(2)$ whose
generators satisfy the commutation relations for angular momentum
(in units of $\hbar = 1$)
\begin{equation}
[S_a, S_b] = i \epsilon_{abc} S_c, \label{spin al}
\end{equation}
where $\epsilon_{abc}$ is the Levi-Civita tensor, taking $+1 (-1)$
for even (odd) permutations of 1, 2, 3 and vanishing otherwise.
The Lie algebra $su(2)$ can be used to find the evolution operator
for the spin-1/2 particle system \cite{shadwick}.

On the other hand, a harmonic oscillator has the Lie algebra
$su(1,1)$, which is isomorphic to $so(2,1)$ \cite{wybourne}. The
generators of $su(1,1)$ produce the spectrum of a number of
quantum systems. These generators satisfy the commutation
relations
\begin{equation}
[K_1, K_2] = - i K_3, \quad [K_2, K_3] = i K_1, \quad [K_3, K_1] =
i K_2.
\end{equation}
We may introduce the position representation of $su(1,1)$ as
\begin{eqnarray}
K_1 = \frac{1}{4} \Bigl(\frac{\partial^2}{\partial q^2} + q^2
\Bigr), \quad K_2 = - \frac{i}{2} \Bigl(q \frac{\partial}{\partial
q} + \frac{1}{2} \Bigr), \quad K_3 = \frac{1}{4}
\Bigl(\frac{\partial^2}{\partial q^2} - q^2 \Bigr).
\end{eqnarray}
In the above definition, we scaled $y = 2 q$ as in Ref.
\cite{wybourne}. The Lie algebra $su(1,1)$ has also been used to
find the quantum states of an oscillator \cite{gerry}.

The compact Lie algebra $su(2)$ is isomorphic to $so(3)$, $(su(2)
\approx so(3))$, whereas the noncompact Lie algebra $su(1,1)$ is
isomorphic to $so(2,1)$, $(su(1,1) \approx so(2,1))$
\cite{wybourne}. In fact, the compact group $SO(3)$ generated by
$so(3)$ holds $x_1^2 + x_2^2 + x_3^2$ invariant whereas the
noncompact group $SO(2,1)$ generated by $so(2,1)$ holds $x_1^2 +
x_2^2 - x_3^2$ invariant. Though the spin generators of $su(2)$
satisfy the commutation relations for an angular momentum
operator, many years ago Schwinger introduced the two-mode bosonic
representation for $su(2)$ \cite{schwinger}. It is, thus, tempting
to exploit the connection between the spin particle and the
oscillator. As $su(2)$ is compact but $su(1,1)$ is noncompact, the
connection between these two algebras requires a complexification
of generators. This can be achieved by introducing a one-mode
bosonic representation of $su(2)$, $S_x = - K_3$, $S_y = - i K_1$,
and $S_z = i K_2$. The generators $S_y$ and $S_z$ are
anti-Hermitian because $K_1$ and $K_2$ are Hermitian operators.
Nevertheless, it is likely that the oscillator may be used to find
the evolution of the spin particle just as the evolution operator
of an oscillator can be found from a spin-1/2 system
\cite{shadwick}.

The density operator of a physical system is an important tool for
understanding the statistical properties of a system. The density
operator, $\rho (t)$, provides an ensemble average, ${\rm Tr}
[\rho (t) A]$, for every physical observable $A$. In this paper,
we introduce a novel method to solve the homogeneous Bloch
equation from the connection with the time-dependent oscillator
that uses the one-mode bosonic representation of $su(2)$ and find
a class of density operators for a spin-1/2 particle in
time-dependent magnetic fields. A time-dependent oscillator is
known to possess invariant operators, or density operators, that
satisfy the Liouville equation \cite{lewis}. We use some known
invariant operators for the oscillator to find the density
operators for the spin particle, for instance, in constant and/or
oscillating magnetic fields.

The organization of this paper is as follows: In Sec. II, we
briefly discuss the density operator for a spin-1/2 particle in a
time-dependent magnetic field. The density operator satisfying the
Liouville equation is determined by a solution of the Bloch or
Landau-Lifshitz equation. In Sec. III, we introduce a one-mode
bosonic representation of $su(2)$ and find the oscillator that
corresponds to the spin-1/2 particle. Further, using an invariant
operator for the oscillator, we find a density operator of the
spin particle through a connection of Lie algebras $su(1,1)$ and
$su(2)$. In Sec. IV, we compare our method with the evolution
operator and the complex parameter method. In Sec. V, we apply the
method to find density operators in constant and/or oscillating
magnetic fields and compare them with the well-known results.

\section{Spin-1/2 Particle in a Magnetic Field}

We consider a spin-1/2 particle under the influence of an external
magnetic field. For the sake of simplicity, we set $\gamma = g
\mu_B = 1$, absorbing it into the magnetic field, where $g$ and
$\mu_B$ are the Land\'{e} g-factor and the Bohr magneton,
respectively, of the spin particle. Then, the spin particle has
the Hamiltonian
\begin{equation}
{\cal H} (t) = - \sum_{a = x, y, z} B_a (t) S_a, \label{spin0}
\end{equation}
where $S_a$ is the $a$th spin component. The spin particle obeys
the time-dependent Schr\"{o}dinger equation for a spinor state:
\begin{equation}
i  \frac{\partial}{\partial t}\vert \Psi , t \rangle = {\cal
H} (t) \vert \Psi , t \rangle.
\end{equation}

It is well understood that any instantaneous eigenstate of a
time-dependent Hamiltonian is not, in general, an exact quantum
state. Nor does the evolution operator follow $U(t) = \exp[-i \int
{\cal H} (t) dt]$ because $[{\cal H} (t'), {\cal H}(t) ] \neq 0 $
for $t' \neq t$. We may directly solve either the Heisenberg or
Schr\"{o}dinger equation. However, instead of solving the
Heisenberg equation, in this paper, we use the invariant operator
method introduced by Lewis and Riesenfeld \cite{lewis}. In fact,
an invariant or density operator satisfying the Liouville equation
\begin{equation}
i  \frac{\partial {\cal I}(t)}{\partial t} + [ {\cal I} (t),{\cal
H}(t)] = 0 \label{leq}
\end{equation}
leads to a general quantum state of the form
\begin{eqnarray}
\vert \Psi  , t \rangle = \sum_{n} c_n e^{i \int \langle n, t
\vert i \frac{\partial}{\partial t} - {\cal H} \vert n, t \rangle}
\vert n, t \rangle, \label{ex st}
\end{eqnarray}
where
\begin{eqnarray}
{\cal I} (t) \vert n, t \rangle = \lambda_n \vert n, t \rangle.
\label{lr}
\end{eqnarray}

For the spin 1/2-particle with $su(2)$, one may find a density
operator of the form \cite{mizrahi}
\begin{equation}
{\cal I} (t) = - \sum_{a = x, y, z} M_{a} (t) S_a, \label{spin
inv}
\end{equation}
where the vector ${\bf M}$ with components $M_a$ satisfies the
Bloch or Landau-Lifshitz equation
\begin{eqnarray}
\frac{d {\bf M}}{dt} + {\bf M} \times {\bf B} = 0. \label{ll eq}
\end{eqnarray}
Note that ${\bf M}^2$ is a constant of the motion. There have been
some attempts to find the evolution operator or density operator
for a spin particle in magnetic fields \cite{mizrahi,lai,ho-kim}.

\section{Connection with an Oscillator}

We now introduce a one-mode bosonic representation for $su(2)$:
\begin{equation}
S_x = \frac{1}{4} (p^2 + q^2), \quad S_y = \frac{i}{4} (p^2 -
q^2), \quad S_z = \frac{i}{4} (pq + qp), \label{osc al}
\end{equation}
where the standard commutation relation $ [q, p] = i $ holds. Note
that our one-mode bosonic representation of $su(2)$ consists of
$S_x = - K_3$, $S_y = - i K_1$, and $S_z = i K_2$. Therefore, the
generators $S_y$ and $S_z$ are anti-Hermitian because $K_1$ and
$K_2$ are Hermitian operators whereas $S_x$ and $K_3$ are both
Hermitian. The non-Hermitian nature of the one-mode representation
follows from the fact that $su(2)$ is compact whereas $su(1,1)$ is
noncompact. Though the bosonic representation in Eq. (\ref{osc
al}) indeed satisfies the spin angular momentum relation in Eq.
(\ref{spin al}), the relation $S_a^2 = I/4$, however, does not
hold for the bosonic representation because the oscillator has an
infinite number of states. The oscillator Hamiltonian
corresponding to the spin Hamiltonian in Eq. (\ref{spin0}) takes
the form
\begin{equation}
H (t) = B_+ (t) \frac{p^2}{2} + B_+^* (t) \frac{q^2}{2} +
\frac{iB_z (t)}{2} \frac{pq + qp}{2}, \label{osc ham}
\end{equation}
with
\begin{equation}
B_+ (t) = \frac{1}{2} (B_x + iB_y), \quad B_- = B_+^*.
\end{equation}
A passing remark is that the Hamiltonian in Eq. (\ref{osc ham}) is
not a Hermitian operator because the $S_a$ are constructed as
non-Hermitian. This means that the spin directions of $S_a$ do not
pertain to a real physical space, but to directions in a complex
space. Nevertheless, the most important point is that the
algebraic relations hold whatever representation one may use. Any
physical interpretation should given to real operators that are
obtained from algebraic relations in the end. Though invariant
under the parity operator, $q \rightarrow -q$ and $p \rightarrow -
p$, the Hamiltonian does not respect time-reversal symmetry, $p
\rightarrow - p$ and $i \rightarrow - i$, and thus cannot be an
example for ${\cal P T}$-invariant quantum mechanics
\cite{bender}.

From now on, we shall work on the non-Hermitian oscillator in Eq.
(\ref{osc ham}). The spin particle or oscillator with $B_+ = 0$ is
trivial, so this case will not be considered anymore. In the case
of $B_+ \neq 0$, the oscillator in Eq. (\ref{osc ham}) may have an
invariant operator of the form \cite{kim-page}
\begin{eqnarray}
a (t) = - i \Biggl[ u (t) p - \frac{1}{B_+} \Biggl( \dot{u} (t) -
i \frac{B_z}{2} u(t) \Biggr)q \Biggr].
\end{eqnarray}
In fact, $a(t)$ satisfies the Liouville equation
\begin{equation}
i \frac{\partial a (t)}{\partial t} + [ a (t), H(t)] = 0 \label{ca
eq}
\end{equation}
when $u$ is a solution to the auxiliary equation
\begin{equation}
\frac{d}{dt} \Biggl(\frac{\dot{u}}{B_+} \Biggr) +
\Biggl[\frac{{\bf B}^2}{4} - B_+ \frac{d}{dt} \Biggl(\frac{i
B_z/2}{B_+} \Biggr) \Biggr] \Biggl(\frac{u}{B_+} \Biggr) = 0,
\label{aux eq}
\end{equation}
or written as $u = B_+^{1/2} v$, $v$ satisfies the equation in
canonical form
\begin{equation}
\ddot v + \Biggl[\frac{{\bf B}^2}{4} - B_+ \frac{d}{dt}
\Biggl(\frac{i B_z/2}{B_+} \Biggr) - \frac{3}{4}
\Biggl(\frac{\dot{B}_+}{B_+} \Biggr)^2 + \frac{1}{2}
\Biggl(\frac{\ddot{B}_+}{B_+} \Biggr) \Biggr] v = 0.
\end{equation}
A second independent solution to Eq. (\ref{aux eq}) may lead to
another invariant operator. However, the complex conjugate $u^*$
cannot, in general, be a solution to Eq. (\ref{aux eq}) due to the
complex coefficient $B_+$. Thus, $a^{\dagger}$ is not another
invariant operator, in contrast with the Hermitian oscillator
case.

Our stratagem to find the density operators for the spin particle
is first to find invariant operators which are quadratic in $q$
and $p$ and then to use the inverse relations in Eq. (9). For
instance, $a^2$ may lead, using Eq. (9), to the density operator
\begin{equation}
{\cal I} (t) = - 2 \Biggl[ \Biggl\{ u^2 + \Biggl( \frac{\dot{u} -
i \frac{B_z}{2} u}{B_+} \Biggr)^2 \Biggr\} S_x - i \Biggl\{ u^2 -
\Biggl( \frac{\dot{u} - i \frac{B_z}{2} u}{B_+} \Biggr)^2 \Biggr\}
S_y + 2 i u \Biggl(\frac{\dot{u} - i \frac{B_z}{2} u}{B_+} \Biggr)
S_z \Biggr]. \label{1st inv}
\end{equation}
Therefore, we obtain the density operator in Eq. (\ref{spin inv})
with the complex components
\begin{eqnarray}
{\cal M}_{x} &=& - 2 \Biggl\{ u^2 + \Biggl( \frac{\dot{u}
- i \frac{B_z}{2} u}{B_+} \Biggr)^2 \Biggr\}, \nonumber\\
{\cal M}_{y} &=& 2 i \Biggl\{ u^2 - \Biggl( \frac{\dot{u}
- i \frac{B_z}{2} u}{B_+} \Biggr)^2 \Biggr\}, \nonumber\\
{\cal M}_{z} &=&  - 4 i u \Biggl( \frac{\dot{u} - i \frac{B_z}{2}
u}{B_+} \Biggr). \label{com}
\end{eqnarray}
A direct calculation shows that the components in Eq. (\ref{com})
satisfy Eq. (\ref{ll eq}) for the density operator. Furthermore,
the complex ${\cal M}$ satisfies the constraint equation
\begin{equation}
{\cal M}_{x}^2 + {\cal M}_{y}^2 + {\cal M}_{z}^2 = 0. \label{sq}
\end{equation}
Note that Eq. (\ref{sq}) is a consistent condition because ${\cal
M}^2$ is a constant of the motion for Eq. (\ref{ll eq}), so the
eigenvalue of the invariant operator, here $\lambda^2 = {\cal
M}^2$, should be constant. The real part $M_{(r)}$ and the
imaginary part $M_{(i)}$ of ${\cal M}$ individually satisfy Eq.
(\ref{ll eq}), a linear equation. We, thus, have two real density
operators:
\begin{equation}
{\cal I}_{(r)} = - \sum_{a=x, y, z} M_{(r)a} S_a, \quad {\cal
I}_{(i)} = - \sum_{a=x, y, z} M_{(i)a} S_a.
\end{equation}
Note that the real and the imaginary parts have the same
magnitude, but they are orthogonal to each other
\begin{equation}
{\bf M}_{(r)}^2 = {\bf M}_{(i)}^2, \quad {\bf M}_{(r)} \cdot {\bf
M}_{(i)} = 0.
\end{equation}

\section{Comparison with Other Methods}

We shall compare the method in Sec. III with other methods: the
evolution operator and the method recently introduced by Kobayashi
\cite{kobayashi1,kobayashi2}. The evolution operator is
determined, for instance, by Eq. (6.23a) of Ref. \cite{gilmore}.
The Schr\"{o}dinger equation leads the evolution operator $U$,
\begin{eqnarray}
U = \left( \matrix{ v& i w \cr i w^* & v^* \cr} \right), \quad v^*
v + w^* w = 1,
\end{eqnarray}
to satisfy the equation
\begin{equation}
\frac{d}{dt} \Biggl(\frac{\dot{x}}{B_-} \Biggr) +
\Biggl[\frac{{\bf B}^2}{4} + B_- \frac{d}{dt} \Biggl(\frac{i
B_z/2}{B_-} \Biggr) \Biggr] \Biggl( \frac{x}{B_-} \Biggr) = 0,
\quad (x = v, w). \label{ev eq}
\end{equation}
Note that Eq. (\ref{ev eq}) is the complex conjugate of Eq.
(\ref{aux eq}), thus implying $x = u^*$. In this sense, Eq.
(\ref{aux eq}) will determine not only the evolution operator but
also the density matrix, as was shown in Sec. III.

Kobaysahi recently introduced two methods for solving the Bloch
equation. In his first method, the Bloch equation is transformed
to the rotating reference frame \cite{kobayashi1}. This idea of a
rotating frame is in essence similar to using the time-dependent
creation and annihilation operators in the Fock space of
time-dependent harmonic oscillators \cite{kim-page,kim-sq}. These
operators are chosen to satisfy the Liouville equation, Eq.
(\ref{ca eq}) or Eq. (\ref{leq}), so the eigenvalue of the number
operator $\hat{N} (t) = \hat{a}^{\dagger} (t) \hat{a} (t)$ is a
constant of the motion, and the exact quantum state in Eq.
(\ref{ex st}) is an instantaneous eigenstate of the number
operator up to a time-dependent phase factor.

On the other hand, in his second method Kobayashi introduced
complex parameters for the magnetization \cite{kobayashi2}. As the
magnitude $|{\bf M}|$ of the magnetization vector ${\bf M}$ is a
constant of the motion, the magnetization vector can be
normalized:
\begin{equation}
{\bf m} = \frac{\bf M}{|{\bf M}|}, \quad m_x^2 + m_y^2 + m_z^2 =
1.
\end{equation}
He then introduced two complex parameters:
\begin{eqnarray}
\xi &=& \frac{m_x + i m_y}{1 - m_z} = \frac{1 + m_z}{m_x - i m_y},
\nonumber\\
\eta &=& - \frac{1 - m_z}{m_x - i m_y} = - \frac{m_x + i m_y}{1 +
m_z}.
\end{eqnarray}
The Bloch equation, Eq. (\ref{ll eq}), in his notation reading as
\begin{equation}
\frac{d {\bf m}}{dt} + \gamma {\bf B} \times {\bf m} = 0,
\label{bloch}
\end{equation}
leads to the Riccati equation
\begin{equation}
\dot{\xi} = \frac{\gamma}{2} (B_y + iB_x) \xi^2 - i \gamma B_z \xi
+ \frac{\gamma}{2} (B_y - i B_x).
\end{equation}
There is a similar Riccati equation for $\eta$. In our notation,
it is written as
\begin{equation}
\dot{\xi} = i \gamma B_- \xi^2 - i \gamma B_z \xi - i \gamma B_+.
\label{ri eq}
\end{equation}
We linearize the Ricatti equation in Eq. (\ref{ri eq}) by
introducing a new variable
\begin{equation}
\xi = \frac{i}{\gamma B_-} \Bigl(\frac{\dot{z}}{z} - i
\frac{\gamma}{2} B_z \Bigr).
\end{equation}
Finally, we can obtain
\begin{equation}
\frac{d}{dt} \Biggl(\frac{\dot{z}}{\gamma B_-} \Biggr) +
\Biggl[\gamma^2 \frac{{\bf B}^2}{4} - \gamma B_- \frac{d}{dt}
\Biggl(\frac{i \gamma B_z/2}{\gamma B_-} \Biggr) \Biggr] \Biggl(
\frac{z}{\gamma B_-} \Biggr) = 0. \label{new eq}
\end{equation}
Note that Eq. (\ref{new eq}) is the same as Eq. (\ref{ev eq}) for
the evolution operator because we scaled $\gamma$ as $\gamma = 1$
in Eq. (\ref{ll eq}) and ${\bf B} \leftrightarrow - {\bf B}$. This
means that our method based on the connection between a spin-1/2
particle in time-dependent magnetic fields and a time-dependent
oscillator with time-dependent mass and/or frequency is equivalent
to the complex parameter method of Kobayashi. Hence, our method
provides a group theoretical foundation for Kobayashi's method.

\section{Exact Density Operators in Constant and/or Oscillating Magnetic Fields}

We now apply our method to find the class of density operators in
Eq.  (\ref{1st inv}) for a spin particle in magnetic fields. To
show the nontrivial nature of the density operator, we find the
density operator in the trivial case of constant and/or
oscillating magnetic fields and compare it with the known result.

\subsection{Constant Field}

First, we directly solve Eq. (\ref{ll eq}) for the density
operator. In the case of a constant field, ${\bf M}$ parallel to
${\bf B}_0$ satisfies Eq. (\ref{ll eq}). The density operator with
the components
\begin{equation}
M_x = M_0 B_{0x}, \quad M_y = M_0 B_{0y}, \quad M_z = M_0 B_{0z},
\label{ll case1}
\end{equation}
where $M_0$ is an arbitrary constant, is proportional to the
Hamiltonian itself. In the coordinate system with the
$z$-direction along ${\bf B}_0$, a second solution may be found:
\begin{equation}
M_x = M_{0 \perp} \cos (B_0t + \varphi), \quad M_y = - M_{0 \perp}
\sin (B_0t + \varphi), \quad M_z = M_{0z}, \label{ll case2}
\end{equation}
where $M_{0 \perp}$, $M_{0z}$, and $\varphi$ are constants that
are determined by the initial conditions. The vector ${\bf M}$
rotates around the magnetic field due to the torque ${\bf \tau} =
{\bf M} \times {\bf B}_0$.

Second, using the connection with the oscillator, we may find the
density operator given in Eq. (\ref{com}). We look for a solution
to Eq. (\ref{aux eq}) of the form
\begin{equation}
u_+ (t) = B_{0+} M_0^{1/2} e^{i (B_0 t + \varphi)/2}
\end{equation}
for a positive constant $M_0$ and a phase constant $\varphi$. The
complex components of the density operator are given by
\begin{eqnarray}
{\cal M}_{x} &=& - \frac{1}{2} M_0 \Bigl[(B_{0x} + i B_{0y})^2
- (B_0 - B_{0z})^2 \Bigr]  e^{i (B_0 t + \varphi)}, \nonumber\\
{\cal M}_{y} &=& \frac{i}{2} M_0 \Bigl[(B_{0x} + i B_{0y})^2
+ (B_0 - B_{0z})^2 \Bigr] e^{i (B_0 t + \varphi)}, \nonumber\\
{\cal M}_{z} &=&  M_0 (B_{0x} + i B_{0y}) (B_0 - B_{0z}) e^{i (B_0
t + \varphi)}.
\end{eqnarray}
Finally, we obtain a real density operator
\begin{equation}
{\cal I}_{(r)} (t) = - \sum_{a = 1}^3 M_{(r) a} S_a, \label{model
inv}
\end{equation}
where
\begin{eqnarray}
M_{(r) x} &=& - \frac{1}{2} M_0 \Bigl[B_{0x}^2 - B_{0y}^2 - (B_0 -
B_{0z})^2 \Bigr] \cos (B_0 t +
\varphi) +  M_0 B_{0x} B_{0y} \sin (B_0 t + \varphi), \nonumber\\
M_{(r) y} &=& - \frac{1}{2} M_0 \Bigl[B_{0x}^2 - B_{0y}^2 + (B_0 -
B_{0z})^2 \Bigr] \sin (B_0 t +
\varphi) -  M_0 B_{0x} B_{0y} \cos (B_0 t + \varphi), \nonumber\\
M_{(r) z} &=&  M_0 (B_0 - B_{0z}) \Bigl[ B_{0x}  \cos (B_0 t +
\varphi) - B_{0y} \sin (B_0 t + \varphi) \Bigr]. \label{model}
\end{eqnarray}

A few remarks are in order. Note that the components in Eq.
(\ref{model}) satisfy Eq. (\ref{ll eq}) or the Liouville equation.
The two eigenvalues of the density operator in Eq. (\ref{model
inv}) are
\begin{equation}
\lambda_{\pm} = \mp \frac{1}{2} M_{(r)} = \mp \frac{1}{2}  M_0
B_0(B_0 - B_{0z}).
\end{equation}
The imaginary components simply given by
\begin{equation}
{\bf M}_{(i)} = - \frac{d}{dt} {\bf M}_{(r)}
\end{equation}
also satisfy Eq. (\ref{ll eq}) because the time derivative of a
solution is another solution. Another solution to Eq. (\ref{aux
eq}),
\begin{equation}
u_- (t) = B_{0+} M_0^{1/2} e^{- i (B_0 t + \varphi)/2},
\end{equation}
leads to the components in Eq. (\ref{model}) for another density
operator now with $B_0$ replaced by $-B_0$ and $(B_0 t + \varphi)$
by $-( B_0 t + \varphi)$.

\subsection{Oscillating Field}

We turn to a spin particle in constant and oscillating magnetic
fields
\begin{equation}
B_+ = B_{0 +} e^{ - i \omega t}, \quad B_z = B_{0 z},
\end{equation}
where $B_x = 2 B_{0+} \cos (\omega t)$ and $B_y = - 2 B_{0+} \sin
(\omega t)$.  The oscillating magnetic field provides a resonance
of the spin particle in the constant field $B_{0z}$. By directly
solving Eq. (\ref{ll eq}), we may find the density operator with
the components \cite{bloch}
\begin{eqnarray}
M_{x} = \frac{2 B_{0+} M_{0z}}{ \omega + B_{0z}} \cos (\omega t),
\quad M_{y} = - \frac{2 B_{0+} M_{0z}}{ \omega + B_{0z}} \sin
(\omega t), \quad M_{z} = M_{0z}, \label{ll case3}
\end{eqnarray}
where $M_{0z}$ is a constant.

We now use the connection with the oscillator to find the density
operator. Even the general case of $B_x = B_{0x} \cos (\omega t) +
B_{0y} \sin(\omega t)$ and $B_y = - B_{0x} \sin (\omega t) +B_{0y}
\cos(\omega t)$ has a solution to Eq. (\ref{aux eq}) of the form
\begin{eqnarray}
u_+ (t) = B_{0 +} M_0^{1/2} e^{i [( \Omega - \omega) t + \varphi
]/2}, \quad \Omega = \sqrt{ {\bf B}_{0}^2 + 2 \omega B_{0z} +
\omega^2}.
\end{eqnarray}
We then have the complex components
\begin{eqnarray}
{\cal M}_{x} &=& - \frac{1}{2} M_0 \Bigl[(B_{0x} + i B_{0y})^2 e^{
- i \omega t}
- (\Omega - \omega  - B_{0z})^2 e^{ i \omega t } \Bigr] e^{ i (\Omega t + \varphi) }, \nonumber\\
{\cal M}_{y} &=& \frac{i}{2} M_0 \Bigl[(B_{0x} + i B_{0y})^2 e^{ -
i \omega t}
+ (\Omega - \omega  - B_{0z})^2 e^{ i \omega t } \Bigr] e^{ i (\Omega t + \varphi) }, \nonumber\\
{\cal M}_{z} &=&  M_0 (B_{0x} + i B_{0y}) (\Omega - \omega -
B_{0z}) e^{i ( \Omega t + \varphi)}. \label{osc b}
\end{eqnarray}
From  Eq. (\ref{osc b}), we finally obtain the density operator
with real components
\begin{eqnarray}
M_{(r) x} &=& - \frac{1}{2} M_0 \Bigl[\Bigl\{ B_{0x}^2 - B_{0y}^2
- (\Omega - \omega - B_{0z})^2 \Bigr\} \cos (\omega t) + 2 B_{0 x}
B_{0 y} \sin (\omega t) \Bigr]
\cos (\Omega t + \varphi) \nonumber\\
&& - \frac{1}{2} M_0 \Bigl[\Bigl\{B_{0x}^2 - B_{0y}^2 + (\Omega -
\omega - B_{0z})^2 \Bigr\} \sin (\omega t) - 2 B_{0 x} B_{0 y}
\cos (\omega t) \Bigr]
\sin (\Omega t + \varphi), \nonumber\\
M_{(r) y} &=& - \frac{1}{2} M_0 \Bigl[\Bigl\{ B_{0x}^2 - B_{0y}^2
+ (\Omega - \omega - B_{0z})^2 \Bigr\} \cos (\omega t) + 2 B_{0 x}
B_{0 y} \sin (\omega t) \Bigr]
\sin (\Omega t + \varphi) \nonumber\\
&& + \frac{1}{2} M_0 \Bigl[\Bigl\{B_{0x}^2 - B_{0y}^2 + (\Omega -
\omega - B_{0z})^2 \Bigr\} \sin (\omega t) - 2 B_{0 x} B_{0 y}
\cos (\omega t) \Bigr]
\cos (\Omega t + \varphi),  \nonumber\\
M_{(r) z} &=&  M_0 (\Omega - \omega - B_{0z}) \Bigl[ B_{0x}  \cos
(\Omega t + \varphi) -  B_{0 y} \sin (\Omega t + \varphi) \Bigr].
\label{model2}
\end{eqnarray}
We, thus, find the general time-dependent density operator. Note
that in the limit of $\omega = 0$ we recover the result in Eq.
(\ref{model}) for a constant field. Using an independent solution
\begin{equation}
u_- (t) = B_{0 +} M_0^{1/2} e^{- i [( \Omega + \omega) t + \varphi
]/2},
\end{equation}
we can find another density operator obtained by replacing
$\Omega$ by $- \Omega$ and $(\Omega t + \varphi)$ by $-(\Omega t +
\varphi)$.

It would be interesting to compare the solutions Eqs.
(\ref{model}) and (\ref{model2}) with those in Sec. III of Ref.
\cite{kobayashi1}. However, our procedure to get solutions is
simpler than that in Refs. \cite{kobayashi1,kobayashi2}.

\section{Conclusion}

A spin-1/2 particle has a compact Lie algebra $su(2)$, which can
be connected through complexification with the oscillator algebra
$su(1,1)$. We introduced a non-Hermitian one-mode bosonic
representation (\ref{osc al}) for $su(2)$. Using the bosonic
representation, we found the oscillator in Eq. (\ref{osc ham})
corresponding to the spin particle in an external magnetic field.
The spin particle in the time-dependent magnetic field corresponds
to a time-dependent oscillator. It is well known that the
time-dependent oscillator has invariant operators which provide
exact quantum states up to time-dependent phase factors. A caveat
is that the bosonic representation for $su(2)$ is non-Hermitian
and, as a consequence, the corresponding oscillator has, in
general, a non-Hermitian Hamiltonian. However, this connection
still provides us with a novel method for finding the nontrivial
density operators for the spin-1/2 particle, even for the
well-known case of constant and/or oscillating magnetic fields.

We have made use of the connection between the spin-1/2 particle
and an oscillator to develop a novel method for finding the
density operator in Eq. (\ref{com}) for the spin particle. The
solution of the auxiliary equation, Eq.  (\ref{aux eq}), leading
to the density operator is found to be the complex conjugate of
the solution of Eq. (\ref{ev eq}) for the evolution operator,
which in turn is the same as the solution of Eq. (\ref{new eq})
for the complex parameters of the magnetization. As illustrative
examples, we applied the method to the spin particle in constant
and/or oscillating magnetic fields. The density operators in Eqs.
(\ref{model}) and (\ref{model2}) have a complicated
time-dependence. In the case of the constant field, the density
operator in Eq. (\ref{ll case1}) or (\ref{ll case2}) is widely
used whereas the density operator in Eq. (\ref{model}) is the most
general time-dependent one, which can also be obtained by using
Kobayashi's method. Similarly, the density operator in Eq.
(\ref{model2}) differs from that in Eq. (\ref{ll case3}) and seems
to be the most general one for the oscillating field. Thus, the
connection with an oscillator provides an effective method for
finding the general density operator for a spin-1/2 particle.

The oscillator with a constant mass and frequency is known to
possess not only time-independent invariant operators but also
time-dependent ones, which lead to squeezed states of
time-independent states \cite{kim-sq}. The physical meanings and
applications of the time-dependent density operator for constant
field and/or oscillating magnetic fields will be addressed in a
future work. The method of this paper can be easily generalized to
spin chain systems. Each spin particle corresponds to an
oscillator; thus, a spin system is equivalent to a system of
coupled oscillators. The invariant operator for the oscillator
chain with time-dependent masses, frequencies, and couplings may
be found in a similar manner \cite{kim-ramos}. It would be
interesting to study the spin chain system in connection with the
oscillator chain system.

\acknowledgments We would like to thank the referee for informing
us Refs. \cite{kobayashi1,kobayashi2}. This work was supported by
the Korea Research Foundation under grant No. KRF-2003-041-C20053.

\end{document}